\documentclass[icps3]{svjour}
\usepackage{graphicx}
\begin{document}
\title{Metal-insulator transition in anisotropic systems}
\titlerunning{Metal-insulator transition in anisotropic systems}
\author{F.\ Milde \and R.\ A.\  R\"{o}mer \and M.\
  Schreiber}                     
\authorrunning{F.\ Milde, R.\ A.\  R\"{o}mer,  M.\ Schreiber}
\institute{Institut f\"{u}r Physik, Technische Universit\"{a}t,
 09107 Chemnitz, Germany}
\maketitle
\begin{abstract}
  We study the three-dimensional Anderson model of localization with
  anisotropic hopping, {\em i.e.}, weakly coupled chains and weakly
  coupled planes. In our extensive numerical study we identify and
  characterize the metal-insulator transition by means of the
  transfer-matrix method and energy level statistics.  Using high
  accuracy data for large system sizes we estimate the critical
  exponent as $\nu=1.6\pm 0.3$.  This is in agreement with its value
  in the isotropic case and in other models of the orthogonal
  universality class.
\end{abstract}
%
\vspace{0.4cm}

Previous studies of Anderson localization \cite{And58} in
three-dimensional (3D) disordered systems with anisotropic hopping
using the transfer-matrix method (TMM)
\cite{LiSEG89,ZamLES96a,PanE94}, multifractal analysis (MFA)
\cite{MilRS97} and energy-level statistics (ELS) \cite{MilR98} show
that an MIT exists even for very strong anisotropy.  
%
%
In Refs.\ \cite{MilRS99a,MilRSU00}, we studied critical properties of
this second-order phase transition with high accuracy. Here we shall
demonstrate the significance of irrelevant scaling exponents for an
accurate determination of the critical disorder $W_c$ and the critical
exponent $\nu$.  Previous highly accurate TMM studies for isotropic
systems of the orthogonal universality class reported
$\nu=1.54\pm0.08$ \cite{Mac94}, $\nu=1.58\pm0.06$ \cite{SleO99a},
$\nu=1.61\pm0.07$, and $\nu=1.54\pm0.03$ \cite{CaiRS99}, whereas for
anisotropic systems of weakly coupled planes $\nu=1.3\pm0.1$ and
$\nu=1.3\pm0.3$ was found \cite{ZamLES96a}. We emphasize that this
variation in theoretical values has its counterpart in the experiments
where a large variation of $\nu$ has been reported with values ranging
from 0.5 \cite{PaaT83} over 1.0 \cite{WafPL99}, 1.3 \cite{StuHLM93},
up to 1.6 \cite{BogSB99}.  Possibly this experimental ``exponent
puzzle'' \cite{StuHLM93} is due to other effects such as
electron-electron interaction \cite{BogSB99}
or sample inhomogeneities \cite{StuHLM93,RosTP94,StuHLM94}. 

A further important aspect of anisotropic hopping besides the question
of universality is the connection to experiments which use uniaxial
stress, tuning disordered Si:P or Si:B systems across the MIT
\cite{PaaT83,WafPL99,StuHLM93,BogSB99}. Applying stress reduces the
distance between the atomic orbitals, the electronic motion
becomes alleviated, and the system changes from insulating to
metallic.  Thus, although the explicit dependence of hopping strength
on stress is material specific and in general not known, it is
reasonable to relate uniaxial stress in a disordered system to an
anisotropic Anderson model with increased hopping between neighboring
planes.

We use the standard Anderson Hamiltonian \cite{And58}
\begin{equation}
  \label{Hand}
  {\bf H} = \sum_{i \ne j} t_{ij} | i \rangle\langle j | + \sum_{i}
  \epsilon_{i} | i \rangle\langle i |
\end{equation}
with orthonormal states $| i \rangle$ corresponding to electrons
located at sites $i=(x,y,z)$ of a regular cubic lattice with periodic
boundary conditions. The potential energies $\epsilon_{i}$ are
independent random numbers drawn uniformly from $[-W/2,W/2]$. The
disorder strength $W$ specifies the amplitude of the fluctuations of
the potential energy. The hopping integrals $t_{ij}$ are non-zero only
for nearest neighbors and depend on the spatial directions, thus
$t_{ij}$ can either be $t_x$, $t_y$ or $t_z$.  We study (i) {\em
  weakly coupled planes} with $t_x=t_y=1$, $t_z=1-\gamma$ and (ii)
{\em weakly coupled chains} with $t_x=t_y=1-\gamma$, $t_z=1$ with
hopping anisotropy $\gamma\in [0,1]$.  For $\gamma=0$ we recover the
isotropic case, $\gamma=1$ corresponds to independent planes or
chains. We note that uniaxial stress would be modeled by weakly
coupled chains after renormalization of the hopping strengths such
that the largest $t$ is set to $1$.

\begin{figure}[t]  
\includegraphics[width=\hsize]{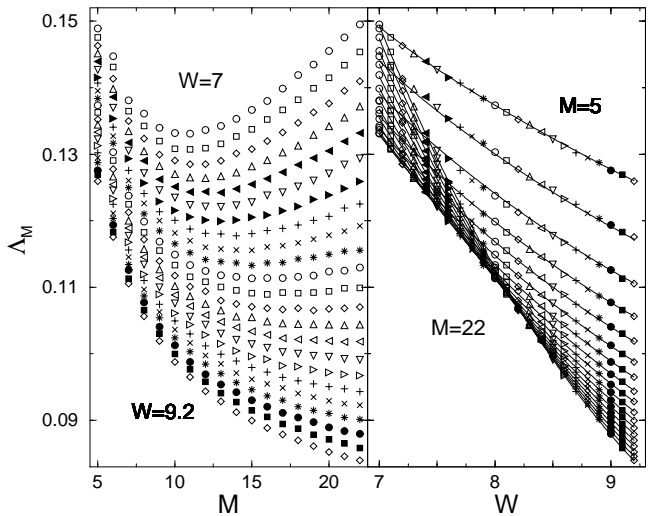}
\caption{\label{fig:tmm-z-0.9} $\Lambda_M$ for coupled planes
  with $\gamma=0.9$ (perpendicular orientation) with relative error
  0.1\%, $W=7,\, 7.1,\, 7.2,\, \cdots,\, 9.2$ and $M^2=5^2,\, 6^2,\, 7^2,\,
  \cdots,\, 22^2$.  The solid lines in the right part are fits to the
  data with $y=2.05\pm 0.08$.}
\label{fig:1}       
\end{figure}

The MIT in the Anderson model of localization is expected to be a
second-order phase transition \cite{BelK94,AbrALR79}. It is
characterized by a divergent correlation length $\xi_\infty(W)\propto
|W-W_c|^{-\nu}$ \cite{KraM93}.  To construct the correlation length of
the {\em infinite} system $\xi_\infty$ from finite size data $\xi_M$
\cite{ZamLES96a,KraM93,PicS81a,MacK81}, the one-parameter scaling
hypothesis \cite{Tho74} $\xi_M=f(M/\xi_\infty)$ is employed.  One
might determine $\nu$ from fitting to $\xi_\infty$ obtained by a FSS
procedure \cite{MacK81}. Better accuracy can be achieved by fitting
directly to the $\xi_M$ data \cite{Mac94,SleO99a,CaiRS99}. We use fit
functions \cite{SleO99a} which include two kinds of corrections to
scaling: (i) nonlinearities of the disorder dependence of the scaling
variable and (ii) an irrelevant scaling variable with exponent $-y$
(cp.\ Fig.\ \ref{fig:tmm-z-0.9}). For the nonlinear fit, we use the
Levenberg-Marquardt method \cite{MilRS99a,SleO99a}. The input data
$\xi_M$ for the FSS procedure are either (a) reduced localization
lengths $\Lambda_M$ obtained by TMM with 0.07\% accuracy and system
widths up to $17\times17$ for, {\em e.g.}, the case of weakly coupled
planes with $\gamma=0.9$ \cite{MilRSU00}; or (b) integrated $\Delta_3$
statistics obtained from highly accurate ELS data (0.2\% to 0.4\%) and
system sizes up to $50^3$ \cite{MilRS99a}.

When applying the TMM to our anisotropic systems, one has to consider
two non-equivalent orientations of the axis of the quasi-1D bar:
parallel and perpendicular to the planes or chains.  The localization
lengths in the perpendicular direction are smaller than in the
parallel direction by a factor of about $1-\gamma$ for coupled planes
and $(1-\gamma)^2$ for chains \cite{ZamLES96a}.  The critical disorder
$W_c$ should not depend on the orientation of the bar
\cite{ZamLES96a}. For strong anisotropies $\gamma\ge0.9$ this is
difficult to verify numerically due to strong finite size effects as
shown in Fig.\ \ref{fig:tmm-z-0.9}.
By computing data for very large system sizes up to $M^2=22^2$
($46^2$) for the case of weakly coupled planes with $\gamma=0.9$
($0.96$) we can show that this finite size effect can be sucessfully
modelled (cp.\ Fig.\ \ref{fig:tmm-z-0.9}) by an irrelevant scaling
exponent and $W_c$ is indeed the same for both orientations.

\begin{figure}[t]  
\includegraphics[width=\hsize]{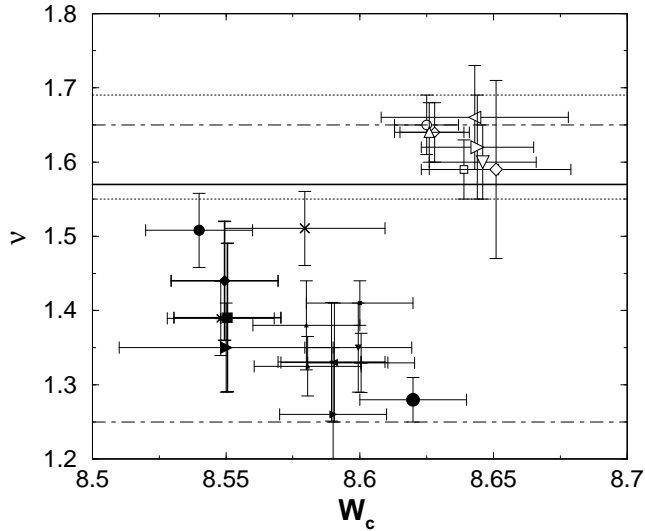}
\caption{
  Results for $W_c$ and $\nu$, for coupled planes with $\gamma=0.9$,
  obtained from FSS of (parallel-direction) TMM data (open symbols)
  and ELS data. The error bars show the $95\%$ confidence intervals.
  The dotted (dashed) lines represent the error bounds for
  $\nu=1.62\pm 0.07$ ($1.45\pm 0.2$) of TMM (ELS). The solid line
  marks the result of \cite{SleO99a}. The goodness of a fit is
  reflected in the size of the symbol.  The $2$ thick error bars mark
  high quality ELS fits for large system sizes. }
\label{fig:nuwc}       
\end{figure}

In Fig.\ \ref{fig:nuwc}, we show fitted values obtained by FSS of TMM
data for different choices of expansion coefficients in the nonlinear
fit procedure. We conclude $\nu=1.62\pm 0.07$ and $W_c= 8.63 \pm
0.02$.
In Fig.\ \ref{fig:nuwc}, we also show the results for FSS of highly
accurate ELS data (0.2\% to 0.4\%) and system sizes up to $N^3=50^3$.
The error estimate is larger and the values of $W_c$ and $\nu$ are
much more scattered than before. Comparing the spreading of the $W_c$
and $\nu$ values with their confidence intervals, the error estimates
appear to be too small. {\em E.g.}, the 95\% confidence intervals of
the smallest and largest $W_c$ value do not overlap. We therefore
estimate $\nu=1.45\pm0.2$ and $W_c= 8.58 \pm 0.06$ \cite{MilRS99a}.

In conclusion, our results confirm the existence of an MIT for
anisotropy $\gamma<1$ for weakly coupled planes found previsouly in
studies using TMM \cite{ZamLES96a}, MFA \cite{MilRS97}, and recently
by ELS \cite{MilR98}.  We have shown that large system sizes, high
accuracies \cite{MilRS99a,MilRSU00} and irrelevant scaling exponents
are necessary to determine the critical behavior reliably. Our results
are in good agreement with other high accuracy TMM studies for the
orthogonal universality class \cite{Mac94,SleO99a,CaiRS99,SleO97}.
These numerical estimates seem to converge towards $\nu\approx 1.6$.

\begin{acknowledgement}
  We are grateful for the support of the DFG through
  Sonderforschungsbereich 393.
\end{acknowledgement}


\end{document}